\documentclass[prx,aps,twocolumn,longbibliography]{revtex4-1} 
\usepackage{graphicx} 
\usepackage{color}
\usepackage{amsmath,amssymb,bm}
\usepackage{lipsum}
\usepackage{hyperref}

\usepackage{color}
\definecolor{bleuf}{rgb}{0,0.44,0.72}
\definecolor{bleuc}{rgb}{0.965,0.965,0.957}
\definecolor{bleu}{rgb}{0,0.24,0.62}
\definecolor{rouge}{rgb}{0.72,0.24,0.14}
\definecolor{ROUGE}{rgb}{0.72,0.24,0.14}
\usepackage{hyperref}
\usepackage{amsmath}
\usepackage{amssymb}

\usepackage{dsfont}

\hypersetup{
    colorlinks=true,                          
    linkcolor=bleuf, 
    citecolor=bleuf, 
    urlcolor=bleuf  } 

\newcommand{\VV}{\bm{\mathcal V}}

\begin{document} 
\title{{\color{black}Flowing active liquids in a pipe: Hysteretic response of polar flocks to external fields}
}
\author{Alexandre Morin}
\affiliation{Univ Lyon, ENS de Lyon, Univ Claude Bernard Lyon 1, CNRS, Laboratoire de Physique, F-69342 Lyon, France}

\author{Denis Bartolo}
\affiliation{Univ Lyon, ENS de Lyon, Univ Claude Bernard Lyon 1, CNRS, Laboratoire de Physique, F-69342 Lyon, France} 
\begin{abstract}
{\color{black}We investigate the response of colloidal flocks to external fields.} We first show that individual colloidal rollers align with external flows  as would a classical spin with  magnetic fields.  Assembling polar active liquids  from colloidal rollers, we  experimentally demonstrate their hysteretic response: confined colloidal flocks can proceed against external flows. We theoretically explain this collective robustness, using an active hydrodynamic description, and show how orientational elasticity and confinement protect the direction of collective motion.  Finally, we exploit the intrinsic bistability of confined active flows to devise self-sustained microfluidic oscillators. 
\end{abstract}
\date{\today} 
\maketitle
For centuries, applying an external pressure difference has remained the only solution  to flow  a liquid in a pipe.  Over the last ten years, by engineering soft materials from self-propelled units we have learned how to drive fluids from within~\cite{Bausch2010,Dogic2012,Goldstein2013,Bricard2013,Sano2015,Plouraboue2015,Granick2016,GoldsteinNJP2016,Dogic2017,Granick_Review}. The generic strategy consists in assembling orientationally ordered  liquids from self-propelled particles~\cite{Marchetti_Review,Vicsek_Review,Toner_Review}. From a fundamental perspective, significant efforts have been devoted to explaining the emergence of collective motion in ensembles of interacting motile bodies, and the flow patterns of the resulting  polar and nematic phases~\cite{Granick_Review,Marchetti_Review,Vicsek_Review,Toner_Review}. However,  we still lack basic understanding of these  non-equilibrium materials.  One of the major questions that remains to be elucidated is the  response of active phases to external fields ~\cite{Vicsek1997,Toner2016}.
The situation is all the more unsatisfactory because, from an applied perspective, the potential of active fluids as smart materials will be chiefly determined by their ability to sustain their spontaneous flows against external perturbations.

Here, combining experiments and theory, we elucidate how confined  active fluids with broken rotational symmetry respond to external fields. Our experiments are based on colloidal rollers self-assembled into  polar flocks, i.e.  active liquids  with orientational order akin to that of a ferromagnet~\cite{Bricard2013,Bricard2015,Morin2017}.
We first demonstrate that isolated colloidal rollers  align their direction of motion  with an external flow as would  classical spins with a magnetic field. In contrast, we establish that the response of  polar liquids  is intrinsically non-linear. When confined  in channels, transverse  confinement and bending elasticity act together to protect the direction of collective motion against  external flows. 
We close this paper showing how the resulting hysteretic relation between the flock velocity and  external flows results in the spontaneous oscillations of confined polar-liquid droplets. 
\section*{Response of motile colloids to external flows: Alignment with external fields}
In our experiments, we exploit the so-called Quincke mechanism to motorize inert colloidal particles and turn them into  self-propelled particles~\cite{Bricard2013}. We recall the motorization principle in Appendix \ref{quincke}, and provide details about the experimental set-up in Appendix \ref{methods}. In brief, we observe the 2D motion of colloidal rollers of diameter $2a=4.8\,\rm \mu m$ confined in $50\,\rm mm\times 16\,mm\times 0.1\,mm$ channels filled with hexadecane oil, as illustrated in Fig.~\ref{Fig1}a. They behave as persistant random walkers moving at a constant speed $v_0=0.98\,\pm0.1\,\rm mm/s$, and having a rotational diffusivity $D=1.5\,\rm s^{-1}$~\cite{Morin2017b}. As a result, the distribution of the roller velocities is isotropic and narrowly peaked on a circle of radius $v_0$, see Fig.~\ref{Fig1}b.  

We  investigate their individual response to external flows by injecting a fresh hexadecane solution at constant flow rate.  Given the aspect ratio of the fluidic channel, the  flow  varies only in the $z$ direction along which it has a Poiseuille profile, Fig.~\ref{Fig1}a. {\color{black}We denote $h$ the  magnitude of the hexadecane flow evaluated at $z=a$ along the $\hat {\mathbf x}$ direction.} 
Over a wide range of $h$, the speed of the rollers is virtually unchanged,  see Figs.~\ref{Fig1}b and~\ref{Fig1}c. Their orientational distribution is however strongly biased.  As seen in Fig.~\ref{Fig1}b, the average roller velocity points in the direction of the flow and the angular fluctuations are reduced upon increasing $h$. More quantitatively this behavior is very well captured by the following equations of motion:
\begin{align}
\partial_t\mathbf r_i(t)&=v_0\hat{\mathbf v}_i,
\label{Eq:r}
\\
\partial_t\theta_i(t)&=-\mu h\sin\theta_i+\sqrt{2D}\xi_i(t),
\label{Eq:theta}
\end{align}
where the $\mathbf r_i(t)$ and $\hat{\mathbf v}_i\equiv(\cos\theta_i(t),\sin\theta_i(t))$ are respectively the  positions and velocity orientations of the colloids. $\mu$ is a constant mobility coefficient and $\xi(t)$ is a Gaussian white noise of unit variance. Eq.~\eqref{Eq:theta} corresponds to the over-damped Langevin dynamics of a classical spin coupled to a constant magnetic field. We henceforth use this magnetic analogy and  define the average roller magnetization as $\bm{\mathcal V}=\langle\mathbf{\hat{v}}_i\rangle_i$.
Eq.~\eqref{Eq:theta} is  readily solved and used to measure the rotational mobility $\mu=0.08\,\rm \mu m^{-1}$ from the magnetization curve in Fig.~\ref{Fig1}d. This value is in excellent agreement with the estimate derived from first principles in~\cite{Bricard2013} and has a sign opposite to that of the colloidal surfers studied in~\cite{Palacci2015}. Isolated rollers align with a flow field as  would uncoupled $XY$ spins with a  magnetic field. 
\section*{Robustness to external fields: Hysteretic response of polar liquids}
\subsection*{Experiments}

In the absence of external flows,  ferromagnetic orientational order  emerges over system-spanning length scales when increasing the roller packing fraction $\rho$ above $10^{-2}$ {\color{rouge}\cite{Bricard2013}}. A homogeneous polar liquid then forms and spontaneously flows through the microchannels as illustrated in Fig.~\ref{Fig2}a and Supplementary video 1.  We now address the  response of this active ferromagnet  to external fields taking advantage of the coupling between the roller velocity and the surrounding fluid flows.
{\color{black} To do so, we assemble a polar liquid confining hundreds of thousands of rollers in a race-track pattern of width $W=0.5\,\rm mm$ (the  area fraction is set to $\rho=0.1$).} Once a homogeneous and stationary polar order is established, we study its longitudinal response  by applying a uniform flow along one of the two straight parts of the channel as sketched in Fig.~\ref{Fig2}a{\color{black}, and detailed in Appendix \ref{methods}}.
For the sake of clarity, we henceforth refer to the hexadecane flow field evaluated at $z=a$ as the external field $\mathbf h=h\hat{\mathbf x}$.  Fig.~\ref{Fig2}b shows that applying a field $\mathbf h$ along the  direction of  $\bm{\mathcal V}$ reduces the transverse velocity fluctuations of collective motion.

 In stark contrast to the individual response,  Fig.~\ref{Fig2}b also shows that collective motion can occur in the direction {\em opposite} to the external field with a high level of ordering. {But this robustness has a limit.} Increasing $|h|$ above $h_{\rm c}=63\,\mu\rm m/s$, the rollers abruptly change their direction of motion and align with $\mathbf h$. As quantitatively demonstrated in Fig.~\ref{Fig2}c, this behavior translates into a strong hysteresis of the magnetization curve  $\VV(h)$ upon cycling the magnitude  of the external field.

{\color{black}We now elucidate the origin  of this collective robustness, or more precisely, what sets the magnitude of the coercive field $h_{\rm c}$ in this active ferromagnet.} We thus focus  on the regime where $\VV\cdot\mathbf h<0$, and leave the discussion of the case where $\VV\cdot\mathbf h>0$ to Appendix \ref{nonlinearresponse}. In an infinite system $h_{\rm c}$ should vanish as the polar-liquid flow stems from the spontaneous  breaking of  a continuous rotational symmetry.  In this active system, however, orientation is intimately coupled to mass transport. Therefore, the homogeneous rotation of the roller velocity is forbidden by the confining boundaries: reversing the direction of the flock requires finite wave-length distortions. As shown in Fig.~\ref{Fig2}d, this picture is supported by the very sharp increase of $h_{\rm c}$ measured when decreasing the channel width. 

To gain a more quantitative insight, we inspect the inner structure of the roller flow field.
Supplementary video 2 and Fig.~\ref{Fig3}a both show that applying a field in the upstream direction causes finite wave-length distortions dominated by bending deformations.
We introduce the Fourier transform of the roller flow field as $\mathbf v(\mathbf q,t)=\langle\mathbf v_i(t)\exp[i\mathbf q\cdot\mathbf r_i(t)]\rangle_i$, and  plot in Fig.~\ref{Fig3}b  the time-averaged spectrum of the bending modes along the $\mathbf q=(q_x,0)$ direction: $B(q_x)=\langle |v_y(q_x,t)|^2\rangle_t$. The bending deformations are dominated by spatial oscillations at a well-defined wave-length $\lambda_{\rm B}$. Fig.~\ref{Fig3}c indicates that confinement sets $\lambda_{\rm B}=2W$, see also Supplementary video 3. Increasing the magnitude of the field strongly increases the amplitude of the bending  oscillations until $|h|=h_{\rm c}$,  Fig.~\ref{Fig3}d.  As exemplified in Supplementary video 4, the bending waves are then destabilized into vortices leading to flow reversal. Subsequently,  the external field  stabilizes a strongly polarized homogeneous polar liquid flowing in the direction of $\mathbf h$.
We stress that the reversal of the flow is completed {\em without} resorting  to local melting. Orientational order is locally preserved regardless of the direction and magnitude of the external field. The weak decrease of the magnetization curve seen in Fig.~\ref{Fig2}c in the negative $\VV\cdot \mathbf h$ region chiefly stems from the constrained oscillations of the spontaneous flow. 
\subsection*{Theory}
{\color{black}We use a hydrodynamic description of the polar liquid  to theoretically account for the bistability of the spontaneous flows and  for the  variations of  the coercive field $h_{\rm c}(W)$ with confinement. The Toner-Tu equations are the equivalents of the Navier-Stokes equations for polar active liquids. They describe the dynamics of the velocity $\mathbf{v} (\mathbf r,t)$ and density fields $\rho (\mathbf r,t)$. In the presence of an external driving field $\mathbf h(\mathbf r,t)$, they take the generic form:
\begin{equation}
\partial_t \rho + \nabla\cdot(\rho \mathbf{v}) =0,
\label{Eq:Masse}
\end{equation}
and
\begin{align}
& \partial_t\mathbf{v} + \lambda_1(\mathbf{v}\cdot\nabla)\mathbf{v}+ \lambda_2(\nabla\cdot\mathbf{v})\mathbf{v} + \lambda_3\nabla(|\mathbf{v}|^2) \label{Eq:TT}\\
& =\alpha\mathbf{v} - \beta|\mathbf{v}|^2\mathbf{v} - \nabla P  \nonumber\\
&+ D_{\rm B} \nabla(\nabla\cdot\mathbf{v}) + D_{\rm T} \nabla^2\mathbf{v} + D_2(\mathbf{v}\cdot\nabla)^2\mathbf{v}+ \mu_h \mathbf{h} \nonumber
\end{align}

These phenomenological equations involve a number of hydrodynamic coefficients which we do not describe here (see e.g.~\cite{Toner_Review} for a comprehensive discussion). The only two parameters relevant to the following discussion are: (i) the convective coefficient $\lambda_1$ which translates the lack of translational invariance of the system: the rollers drag on a fixed substrate defining a specific frame. (ii) $D_{\rm T}$ and $D_2$ which are two elastic constants of the broken symmetry fluid.  $D_{\rm T}$ and $D_2$ both hinder the orientational distortions of the velocity field. 

Describing the flow-reversal dynamics and the underlying spatiotemporal patterns would require solving the  strongly non-linear system given by Eqs.~\eqref{Eq:Masse} and ~\eqref{Eq:TT}. This task  goes far beyond the scope of this article. Here,  we instead exploit our experimental observations to construct a minimal model. As Fig.~\ref{Fig3} indicates that a single bending mode of wavevector $\mathbf q=\pi/W \hat{\mathbf x}$ dominates the deformations of the velocity and density fields, we  make a simplifying  ansatz writing: $\mathbf{v}(x,t) =v_x(t)\mathbf{\hat{x}} +v_y(t)\cos(qx-\omega t)\mathbf{\hat{y}}$, and neglect all contributions from spatial frequencies higher than  $\pi/W$.
We also ignore density fluctuations and restrain our analysis to longitudinal external fields $\mathbf{h} = -h\mathbf{\hat{x}}$. Within this two-mode approximation, Eq.~\eqref{Eq:Masse} is always satisfied and Eq.~\eqref{Eq:TT} reduces to the dynamical system:
\begin{equation}
\partial_t \mathbf{V} = \mathbf{F}(\mathbf{V},h) 
\label{Eq:TT2modes}
\end{equation}
for the amplitudes of the two coupled velocity modes: $\mathbf{V}(t)=(v_x(t),v_y(t))$. The generalized force $\mathbf{F}(\mathbf{V},h) = (F_x,F_y)$ is given by
\begin{align}
& F_x =  \left[ \alpha - \beta \left( v_x^2 +\frac{1}{2}v_y^2 \right) \right] v_x - h \label{Eq:DynSystX}\\
& F_y =  \left[ \left(\alpha - D_{\rm T}q^2\right) - D_2q^2v_x^2 -  \beta \left( v_x^2 +\frac{3}{4}v_y^2 \right) \right] v_y
\label{Eq:DynSystY}
\end{align}
  
To gain more intuition on the physical meaning of the dynamical system, we plot the force field $\mathbf F(v_x,v_y)$ in Fig.~\ref{Fig4} for four values of $h$.  
Anticipating on the comparison with our measurements, we  use the parameter values estimated in~\cite{Bricard2013} and recalled in Appendix~\ref{TonerTu}. 
 Looking for fixed points in the absence of external field, $\mathbf{F}(\mathbf{V},h=0) = 0$, we find five solutions for $\mathbf{V}$ when $\alpha>D_{\rm T} q^2$, see Fig.~\ref{Fig4}a
 \footnote{We  focus on the situation where $\alpha>D_{\rm T}q^2$. In the case of extreme confinement where this condition is not met, flow reversal can occur only via local melting.}.
  The trivial solution $(0,0)$  is obviously unstable and corresponds to a fluid at rest. The four other solutions are given by $(0,v_{y,\pm}) = \pm\left(0,2\sqrt{{(\alpha-D_{\rm T}q^2)}/{(3\beta)}}\right)$ and  $(v_{x,\pm},0) = \pm\left(\sqrt{{\alpha}/{\beta}},0\right)$ as illustrated in Fig.~\ref{Fig4}a. The former are  saddle points while the latter are linearly stable fixed points  corresponding to the two possible homogeneous flows at  speed  $v_0 = \sqrt{{\alpha}/{\beta}}$ along the  $\pm\mathbf{\hat{x}}$ directions. {It is worth noting that the force field focuses the position of the dynamical system along a closed line connecting the four fixed points thereby making the dynamics of $\mathbf V$ almost one dimensional, see Fig.~\ref{Fig4}a.}

The bistability of the flow at finite $h$, viz the existence of a finite coercive field, is understood from the dynamics of the fixed points in  phase space. Upon increasing $h$ the dynamical system, and therefore the active flow, explore three different states  labeled with a star symbol in Figs.~\ref{Fig4}b, \ref{Fig4}c and \ref{Fig4}d:

(i) We start from a polar liquid flowing in the $+\mathbf{\hat{x}}$ direction, and an opposing  external field $\mathbf h=-h\hat{\mathbf x}$, with $h>0$. As sketched in Fig.~\ref{Fig4}b, this state  where  $\mathbf{V}=(v_0-h/(2\alpha),0)$ corresponds to a polar fluid uniformly flowing against the external field. This situation remains stable until $h$ reaches $h_{\rm b} = D_{\rm T}v_0 q^2+{\mathcal O}(D_{\rm T} q^2/\alpha)$, Fig.~\ref{Fig4}b. Above this value, the homogeneous flow is unstable to buckling, and  the stable point $\mathbf{V}=(v_0-h/(2\alpha),0)$ becomes a saddle point. 

(ii)  {Yet, the flow is not reversed.} The system indeed reaches one of the two new stable fixed points with $v_y\neq0$. They both correspond to  homogeneously buckled conformations, see Fig.~\ref{Fig4}c.  This prediction is consistent with the buckled patterns  observed prior to  flow reversal  shown in Fig.~\ref{Fig3} and Supplementary videos 2 and 3. 

(iii) Further increasing $h$, the buckled state approaches the topmost saddle point. The two points eventually merge at a critical value $h_{\rm c}$ corresponding to Fig.~\ref{Fig4}d. The only stable conformation then  corresponds to a situation where $\mathbf{V}=(-v_x,0)$. The flow is reversed and aligns along the direction prescribed by $\mathbf h$. 
 $h_{\rm c}$  defines the value of the coercive field. 
 
 The value of $h_{\rm c}$ is determined analytically by the merging condition between the saddle and the fixed point, see Fig.~\ref{Fig4}d. We find that $h_{\rm c}$ stems from the competition between the external field and all the velocity-alignment terms ($\alpha$ and $D_{\rm T}$):  $h_{\rm c}=(2/\sqrt{243})\alpha v_0\left(1+2D_{\rm T} q^2/\alpha\right)^{3/2}$.   Our model correctly predicts that 
the stability of the flows opposing an external  field is enhanced when  further confining the polar liquid, i.e increasing $q=\pi/W$.  Remarkably, this  simplified picture also provides a reasonable estimate of the magnitude of the coercive field, see Figs.~\ref{Fig2}d, and~\ref{Fig2}e. 

In summary, we have  established the bistability of  polar active fluids. Their 
hysteretic  response  originates from  buckled  flow patterns stabilized by orientational elasticity. We  expect this phenomenology to apply to all confined active fluids with uniaxial orientational order.   Our theory builds on the observation of a single set of buckling modes. Explaining  the pattern-selection process remains,  however, a   significant technical challenge.

\section*{Application: Spontaneous oscillations of polar-liquid droplets}
We close this article  exploiting the intrinsic multistability of polar-liquid flows and demonstrating emergent  functionalities in active microfluidics~\cite{DunkelPNAS,DunkelPRL,DunkelNatComm}.  The existence of a hysteresis loop in the response function provides a very natural design strategy for spontaneous oscillators via the relaxation-oscillation mechanism~\cite{Nekorkin2015}.  Simply put, and having mechanical devices in mind, relaxation oscillations stem from the coupling between a system with a hysteretic ``force-velocity'' relation and a harmonic spring. This minimal design rule is transposed to active fluids by confining them in curved containers, and applying a constant and homogeneous external field $\mathbf h= h\hat{\mathbf x}$. As illustrated in Supplementary video 5, and in the image sequence of Fig.~\ref{Fig5}a, the colloids form  a polar-liquid droplet that spontaneously glides along the confining boundary in an oscillatory fashion. {\color{black}We denote $\alpha$ the polar angle defining the position along the confining disc, and $\mathbf v(\alpha,t)$ the azimuthal component of the velocity field averaged over the radial direction. Fig.~\ref{Fig5}b shows the variations of the velocity field $\mathbf v(\alpha,t)$. The oscillatory dynamics of the polar-liquid droplets is clearly  periodic with well-defined period and amplitude both decreasing with the magnitude of the  stationary external field, Figs.~\ref{Fig5}c, and~\ref{Fig5}d. 

We now explain these  collective oscillations  as the periodic exploration of the four states (i), (ii), (iii) and (iv) along the hysteresis loop {established} in Fig.~\ref{Fig2}c and sketched in \ref{Fig2}e . The key observation is that the droplet follows the boundary of the circular chamber.  The droplet hence experiences a longitudinal  field of magnitude $h_\parallel(\alpha)=h\sin\alpha$ which either favors or hinders its motion. The periodic exploration of the hysteresis loop is  supported by Fig.~\ref{Fig5}e.  Fig.~\ref{Fig5}e shows the distribution $P(v_{\rm CoM},h_\parallel)$, where $v_{\rm CoM}$ is the polar-liquid velocity   evaluated at the center of mass of the droplet $\alpha_{\rm CoM}$. The support of this distribution  corresponds to the rectangular shape of the velocity-field relation measured in Fig.~\ref{Fig2} for a straight channel. The droplet spends most of its time exploring the stable horizontal branches and quickly jumps from one stable conformation to the other along the vertical ones. We can gain more intuition on this oscillatory dynamics  describing the four states one at a time: }

In state (i), the head of the flock is located at $\alpha<0$ and  $\mathbf v(\alpha)\cdot\mathbf h<0$. 
The flock proceeds in the direction opposite to the azimutal component of the external field. The system moves toward the left of the bottom branch of the hysteresis loop, Fig.~\ref{Fig5}e.
As the flock moves toward the negative $\alpha$ direction, the field strength $|h_\parallel|$ increases and reaches the maximal value $h_{\rm c}$ at an angle $-\alpha_{\rm c}$. The system then reaches the left vertical branch of the response curve and  hence becomes unstable, state (ii). The flock bends and reverses its direction to reach the upper branch of the response curve,  state (iii).  When the flock proceeds in  the positive $\alpha$ region, it experiences an increasingly high field in the  direction opposite to its motion. 
As $h_\parallel=h_{\rm c}$ the flock reaches the right vertical branch of the response curve at the maximal angle $+\alpha_{\rm c}$ (state iv), thereby leading the system back to state (i). The hysteresis loop is periodically explored.

{\color{black}This oscillatory motion relates to the conventional relaxation-oscillation picture as follows: the response curve $(h,{\mathcal V})$ plays the role of the force-velocity relation in a mechanical system, and the angle-dependent longitudinal flow plays the role of the harmonic spring.   }
 
\section*{Conclusion}
In conclusion, we have established that colloidal rollers respond to external flows as classical spins to magnetic fields. Assembling active fluids with broken orientational symmetry from these elementary units, we have experimentally demonstrated, and theoretically explained, the hysteretic response of polar-active-fluid flows. We have  shown how confinement and bending elasticity act together to protect emergent flows against external fields. Finally, we have effectively exploited the  bistability of active flows to   engineer active-fluid oscillators with frequency and amplitude set by the geometry of the container.  Together with  the virtually unlimited geometries accessible to microfabrication,  the intrinsic nonlinearity of active flows  offer an effective  framework for the design of emergent microfluidic functions ~\cite{DunkelPNAS,DunkelPRL,DunkelNatComm}. 

\appendix
\section{Motorizing colloidal rollers.}
\label{quincke}
Our experiments are based on colloidal rollers, see~\cite{Bricard2013}.  We turn inert polystyrene colloids of diameter $a=4.8\,\rm\mu m$  into self-propelled bodies by taking advantage of the so-called Quincke electro-rotation mechanism~\cite{Quincke,Taylor69}. Applying an electric field to an insulating body immersed in a conducting fluid results in  a dipolar distribution of its surface charges.  Increasing the magnitude of the electric field, $E_0$, above the Quincke threshold $E_{\rm Q}$  destabilizes  the dipole orientation, which in turn makes a finite angle with the electric field. A net electric torque builds up and competes with viscous friction to power the spontaneous rotation of the colloids at constant angular velocity. As sketched in Fig.~\ref{Fig1}a, the colloids are let to sediment on a flat electrode, rotation is then readily converted into translational motion at constant speed $v_0$ in the direction opposite to the charge dipole.
The direction of motion is randomly chosen and freely diffuses as a result of the spontaneous symmetry breaking of the surface-charge distribution.

\section{ Methods}
\label{methods}
We disperse commercial polystyrene colloids (Thermo Scientific G0500) in a mixture of hexadecane and AOT with concentration $[AOT]=0.13\,\rm mol/L$. We inject this solution into microfluidic chambers made of two electrodes spaced by a $110\,\rm \mu m$-thick scotch tape. The electrodes are glass slides, coated with indium tin oxide (Solems, ITOSOL30, thickness: $80\,\rm nm$). A voltage amplifier (TREK 609E-6) applies a DC electric field between the two electrodes. We image the system with a Nikon AZ100 microscope with a 3.6X magnification and record videos with a CMOS camera (Basler Ace) at framerate up to $380\,\rm Hz$. We use conventional techniques to detect and track all particles~\cite{Grier,Lu2007,Blair}. To confined the colloidal rollers inside  racetracks, we pattern the bottom electrode by mean of photolithography using  a $2\,\rm \mu m$-thick layer of UV photoresist (Microposit S1818) as in~\cite{Morin2017}. {\color{black} The geometry of the microfluidic device is detailed in Fig.~\ref{Fig:experience}.
We study the response of rollers to external field, by  injecting a fresh hexadecane solution at a controlled flow-rate using a high-precision syringe pump (Cetoni neMESYS). Each measurement was done at least 60 seconds after the relaxation of the flow pattern in the main branch of the racetrack. The construction of the hysteresis loop in Fig.~\ref{Fig2}b corresponds to a seven-hour long experiment.}

\section{Non-linear response of the ordered phase: ${\mathcal V}\cdot h>0$}
\label{nonlinearresponse}
We discuss here the strengthening of orientational order when $\VV\cdot\mathbf h>0$. We plot in Fig.~\ref{FigSupp1}a the variations of $\delta {\mathcal V}\equiv |{\mathcal V}(h)-{\mathcal V}(h=0)|$ in this regime. At small $h$,  $\mathcal V$ responds linearly to the external field. However, the increase of $\delta \mathcal V(h)$ becomes sub-linear for field amplitudes as small as $h=3\times10^{-2}v_0$. The simplest possible explanation of this anomalous attenuation is that  $\mathcal V$ is a bounded quantity which is  maximal and equals 1 when all the rollers move along the very same direction. A second and more involved  explanation was put forward in~\cite{Toner2016}. For finite systems, a crossover from  linear response at small $h$ to the anomalous scaling law $\delta \mathcal V(h) \sim h^{1/3}$ was predicted from renormalization group analysis~\cite{Toner2016}. As shown in Fig.~\ref{FigSupp1}a, this scaling law is consistent with our experiments for system sizes ranging from $W=0.175\,\rm mm$ to $W=0.5\,\rm mm$. However, the finite size scaling shown in Fig.~\ref{FigSupp1}b fails to ascertain this explanation. Disentangling the two effects would require operating closer to the transition toward collective motion where the fluctuations of $\mathcal V$ are more prominent. Such a regime cannot be achieved in our experiment due to the strongly first-order nature of the transition toward collective motion.
\section{Hydrodynamic parameters of the roller fluid}
\label{TonerTu}
We recall the estimates of the hydrodynamic parameters relevant to the computation of the coercive field $h_{\rm c}$. Starting from the Stokes and Maxwell equations describing the microscopic dynamics of the  colloids,  
we established in~\cite{Bricard2013} the hydrodynamics of colloidal-roller liquids. The results of this kinetic theory are summarized in Table~\ref{Tab:Parameters}. We determined the value of $\mu_h$ following the same procedure, and found $\mu_h=\frac{1}{2}\mu v_0$,  where $\mu$ is the rotational mobility  measured in Fig.~\ref{Fig1}.
\begin{table}[h]
\centering
\begin{tabular}{|c|c|c|c|c|c|}
\hline
$\alpha$ & $\beta$ &  $D_{\rm T}$ & $D_2$ & $\mu_h$ \\ \hline
$50\,\rm s^{-1}$ & $50\,\rm mm^{-2}s^{1}$ &  $2\times10^{-3} \,\rm mm^2s^{-1}$ & $0$ & $40\,\rm s^{-1}$ \\ \hline
\end{tabular}
\caption{Values of the hydrodynamic coefficients of the colloidal-roller liquid.}
\label{Tab:Parameters}
\end{table}

\begin{acknowledgments}
We acknowledge support from ANR program MiTra and Institut Universitaire de France. We thank  M.C. Marchetti for valuable discussions.\\
\end{acknowledgments}



\begin{thebibliography}{31}%
\makeatletter
\providecommand \@ifxundefined [1]{%
 \@ifx{#1\undefined}
}%
\providecommand \@ifnum [1]{%
 \ifnum #1\expandafter \@firstoftwo
 \else \expandafter \@secondoftwo
 \fi
}%
\providecommand \@ifx [1]{%
 \ifx #1\expandafter \@firstoftwo
 \else \expandafter \@secondoftwo
 \fi
}%
\providecommand \natexlab [1]{#1}%
\providecommand \enquote  [1]{``#1''}%
\providecommand \bibnamefont  [1]{#1}%
\providecommand \bibfnamefont [1]{#1}%
\providecommand \citenamefont [1]{#1}%
\providecommand \href@noop [0]{\@secondoftwo}%
\providecommand \href [0]{\begingroup \@sanitize@url \@href}%
\providecommand \@href[1]{\@@startlink{#1}\@@href}%
\providecommand \@@href[1]{\endgroup#1\@@endlink}%
\providecommand \@sanitize@url [0]{\catcode `\\12\catcode `\$12\catcode
  `\&12\catcode `\#12\catcode `\^12\catcode `\_12\catcode `\%12\relax}%
\providecommand \@@startlink[1]{}%
\providecommand \@@endlink[0]{}%
\providecommand \url  [0]{\begingroup\@sanitize@url \@url }%
\providecommand \@url [1]{\endgroup\@href {#1}{\urlprefix }}%
\providecommand \urlprefix  [0]{URL }%
\providecommand \Eprint [0]{\href }%
\providecommand \doibase [0]{http://dx.doi.org/}%
\providecommand \selectlanguage [0]{\@gobble}%
\providecommand \bibinfo  [0]{\@secondoftwo}%
\providecommand \bibfield  [0]{\@secondoftwo}%
\providecommand \translation [1]{[#1]}%
\providecommand \BibitemOpen [0]{}%
\providecommand \bibitemStop [0]{}%
\providecommand \bibitemNoStop [0]{.\EOS\space}%
\providecommand \EOS [0]{\spacefactor3000\relax}%
\providecommand \BibitemShut  [1]{\csname bibitem#1\endcsname}%
\let\auto@bib@innerbib\@empty
\bibitem [{\citenamefont {Schaller}\ \emph {et~al.}(2010)\citenamefont
  {Schaller}, \citenamefont {Weber}, \citenamefont {Semmrich}, \citenamefont
  {Frey},\ and\ \citenamefont {Bausch}}]{Bausch2010}%
  \BibitemOpen
  \bibfield  {author} {\bibinfo {author} {\bibfnamefont {Volker}\ \bibnamefont
  {Schaller}}, \bibinfo {author} {\bibfnamefont {Christoph}\ \bibnamefont
  {Weber}}, \bibinfo {author} {\bibfnamefont {Christine}\ \bibnamefont
  {Semmrich}}, \bibinfo {author} {\bibfnamefont {Erwin}\ \bibnamefont {Frey}},
  \ and\ \bibinfo {author} {\bibfnamefont {Andreas~R}\ \bibnamefont {Bausch}},\
  }\bibfield  {title} {\enquote {\bibinfo {title} {Polar patterns of driven
  filaments},}\ }\href
  {http://www.nature.com/nature/journal/v467/n7311/abs/nature09312.html}
  {\bibfield  {journal} {\bibinfo  {journal} {Nature}\ }\textbf {\bibinfo
  {volume} {467}},\ \bibinfo {pages} {73} (\bibinfo {year} {2010})}\BibitemShut
  {NoStop}%
\bibitem [{\citenamefont {Sanchez}\ \emph {et~al.}(2012)\citenamefont
  {Sanchez}, \citenamefont {Chen}, \citenamefont {DeCamp}, \citenamefont
  {Heymann},\ and\ \citenamefont {Dogic}}]{Dogic2012}%
  \BibitemOpen
  \bibfield  {author} {\bibinfo {author} {\bibfnamefont {Tim}\ \bibnamefont
  {Sanchez}}, \bibinfo {author} {\bibfnamefont {Daniel T.~N.}\ \bibnamefont
  {Chen}}, \bibinfo {author} {\bibfnamefont {Stephen~J.}\ \bibnamefont
  {DeCamp}}, \bibinfo {author} {\bibfnamefont {Michael}\ \bibnamefont
  {Heymann}}, \ and\ \bibinfo {author} {\bibfnamefont {Zvonimir}\ \bibnamefont
  {Dogic}},\ }\bibfield  {title} {\enquote {\bibinfo {title} {Spontaneous
  motion in hierarchically assembled active matter},}\ }\href
  {http://dx.doi.org/10.1038/nature11591} {\bibfield  {journal} {\bibinfo
  {journal} {Nature}\ }\textbf {\bibinfo {volume} {491}},\ \bibinfo {pages}
  {431--434} (\bibinfo {year} {2012})}\BibitemShut {NoStop}%
\bibitem [{\citenamefont {Wioland}\ \emph {et~al.}(2013)\citenamefont
  {Wioland}, \citenamefont {Woodhouse}, \citenamefont {Dunkel}, \citenamefont
  {Kessler},\ and\ \citenamefont {Goldstein}}]{Goldstein2013}%
  \BibitemOpen
  \bibfield  {author} {\bibinfo {author} {\bibfnamefont {Hugo}\ \bibnamefont
  {Wioland}}, \bibinfo {author} {\bibfnamefont {Francis~G.}\ \bibnamefont
  {Woodhouse}}, \bibinfo {author} {\bibfnamefont {J\"orn}\ \bibnamefont
  {Dunkel}}, \bibinfo {author} {\bibfnamefont {John~O.}\ \bibnamefont
  {Kessler}}, \ and\ \bibinfo {author} {\bibfnamefont {Raymond~E.}\
  \bibnamefont {Goldstein}},\ }\bibfield  {title} {\enquote {\bibinfo {title}
  {Confinement stabilizes a bacterial suspension into a spiral vortex},}\
  }\href {https://link.aps.org/doi/10.1103/PhysRevLett.110.268102} {\bibfield
  {journal} {\bibinfo  {journal} {Phys. Rev. Lett.}\ }\textbf {\bibinfo
  {volume} {110}},\ \bibinfo {pages} {268102} (\bibinfo {year}
  {2013})}\BibitemShut {NoStop}%
\bibitem [{\citenamefont {Bricard}\ \emph {et~al.}(2013)\citenamefont
  {Bricard}, \citenamefont {Caussin}, \citenamefont {Desreumaux}, \citenamefont
  {Dauchot},\ and\ \citenamefont {Bartolo}}]{Bricard2013}%
  \BibitemOpen
  \bibfield  {author} {\bibinfo {author} {\bibfnamefont {Antoine}\ \bibnamefont
  {Bricard}}, \bibinfo {author} {\bibfnamefont {Jean-Baptiste}\ \bibnamefont
  {Caussin}}, \bibinfo {author} {\bibfnamefont {Nicolas}\ \bibnamefont
  {Desreumaux}}, \bibinfo {author} {\bibfnamefont {Olivier}\ \bibnamefont
  {Dauchot}}, \ and\ \bibinfo {author} {\bibfnamefont {Denis}\ \bibnamefont
  {Bartolo}},\ }\bibfield  {title} {\enquote {\bibinfo {title} {Emergence of
  macroscopic directed motion in populations of motile colloids},}\ }\href
  {https://www.nature.com/nature/journal/v503/n7474/full/nature12673.html}
  {\bibfield  {journal} {\bibinfo  {journal} {Nature}\ }\textbf {\bibinfo
  {volume} {503}} (\bibinfo {year} {2013})}\BibitemShut {NoStop}%
\bibitem [{\citenamefont {Nishiguchi}\ and\ \citenamefont
  {Sano}(2015)}]{Sano2015}%
  \BibitemOpen
  \bibfield  {author} {\bibinfo {author} {\bibfnamefont {Daiki}\ \bibnamefont
  {Nishiguchi}}\ and\ \bibinfo {author} {\bibfnamefont {Masaki}\ \bibnamefont
  {Sano}},\ }\bibfield  {title} {\enquote {\bibinfo {title} {Mesoscopic
  turbulence and local order in janus particles self-propelling under an ac
  electric field},}\ }\href {\doibase 10.1103/PhysRevE.92.052309} {\bibfield
  {journal} {\bibinfo  {journal} {Phys. Rev. E}\ }\textbf {\bibinfo {volume}
  {92}},\ \bibinfo {pages} {052309} (\bibinfo {year} {2015})}\BibitemShut
  {NoStop}%
\bibitem [{\citenamefont {Creppy}\ \emph {et~al.}(2015)\citenamefont {Creppy},
  \citenamefont {Praud}, \citenamefont {Druart}, \citenamefont {Kohnke},\ and\
  \citenamefont {Plourabou\'e}}]{Plouraboue2015}%
  \BibitemOpen
  \bibfield  {author} {\bibinfo {author} {\bibfnamefont {Adama}\ \bibnamefont
  {Creppy}}, \bibinfo {author} {\bibfnamefont {Olivier}\ \bibnamefont {Praud}},
  \bibinfo {author} {\bibfnamefont {Xavier}\ \bibnamefont {Druart}}, \bibinfo
  {author} {\bibfnamefont {Philippa~L.}\ \bibnamefont {Kohnke}}, \ and\
  \bibinfo {author} {\bibfnamefont {Franck}\ \bibnamefont {Plourabou\'e}},\
  }\bibfield  {title} {\enquote {\bibinfo {title} {Turbulence of swarming
  sperm},}\ }\href {https://link.aps.org/doi/10.1103/PhysRevE.92.032722}
  {\bibfield  {journal} {\bibinfo  {journal} {Phys. Rev. E}\ }\textbf {\bibinfo
  {volume} {92}},\ \bibinfo {pages} {032722} (\bibinfo {year}
  {2015})}\BibitemShut {NoStop}%
\bibitem [{\citenamefont {Yan}\ \emph {et~al.}(2016)\citenamefont {Yan},
  \citenamefont {Han}, \citenamefont {Zhang}, \citenamefont {Xu}, \citenamefont
  {Luijten},\ and\ \citenamefont {Granick}}]{Granick2016}%
  \BibitemOpen
  \bibfield  {author} {\bibinfo {author} {\bibfnamefont {Jing}\ \bibnamefont
  {Yan}}, \bibinfo {author} {\bibfnamefont {Ming}\ \bibnamefont {Han}},
  \bibinfo {author} {\bibfnamefont {Jie}\ \bibnamefont {Zhang}}, \bibinfo
  {author} {\bibfnamefont {Cong}\ \bibnamefont {Xu}}, \bibinfo {author}
  {\bibfnamefont {Erik}\ \bibnamefont {Luijten}}, \ and\ \bibinfo {author}
  {\bibfnamefont {Steve}\ \bibnamefont {Granick}},\ }\bibfield  {title}
  {\enquote {\bibinfo {title} {{Reconfiguring active particles by electrostatic
  imbalance}},}\ }\href {http://www.nature.com/articles/nmat4696} {\bibfield
  {journal} {\bibinfo  {journal} {Nature Materials}\ }\textbf {\bibinfo
  {volume} {15}},\ \bibinfo {pages} {1095--1099} (\bibinfo {year}
  {2016})}\BibitemShut {NoStop}%
\bibitem [{\citenamefont {Wioland}\ \emph {et~al.}(2016)\citenamefont
  {Wioland}, \citenamefont {Lushi}, \citenamefont {E.},\ and\ \citenamefont
  {Goldstein}}]{GoldsteinNJP2016}%
  \BibitemOpen
  \bibfield  {author} {\bibinfo {author} {\bibfnamefont {H.}~\bibnamefont
  {Wioland}}, \bibinfo {author} {\bibnamefont {Lushi}}, \bibinfo {author}
  {\bibnamefont {E.}}, \ and\ \bibinfo {author} {\bibfnamefont {R.~E.}\
  \bibnamefont {Goldstein}},\ }\bibfield  {title} {\enquote {\bibinfo {title}
  {Directed collective motion of bacteria under channel confinement},}\ }\href
  {http://stacks.iop.org/1367-2630/18/i=7/a=075002} {\bibfield  {journal}
  {\bibinfo  {journal} {New Journal of Physics}\ }\textbf {\bibinfo {volume}
  {18}},\ \bibinfo {pages} {075002} (\bibinfo {year} {2016})}\BibitemShut
  {NoStop}%
\bibitem [{\citenamefont {Wu}\ \emph {et~al.}(2017)\citenamefont {Wu},
  \citenamefont {Hishamunda}, \citenamefont {Chen}, \citenamefont {DeCamp},
  \citenamefont {Chang}, \citenamefont {Fern{\'a}ndez-Nieves}, \citenamefont
  {Fraden},\ and\ \citenamefont {Dogic}}]{Dogic2017}%
  \BibitemOpen
  \bibfield  {author} {\bibinfo {author} {\bibfnamefont {Kun-Ta}\ \bibnamefont
  {Wu}}, \bibinfo {author} {\bibfnamefont {Jean~Bernard}\ \bibnamefont
  {Hishamunda}}, \bibinfo {author} {\bibfnamefont {Daniel T.~N.}\ \bibnamefont
  {Chen}}, \bibinfo {author} {\bibfnamefont {Stephen~J.}\ \bibnamefont
  {DeCamp}}, \bibinfo {author} {\bibfnamefont {Ya-Wen}\ \bibnamefont {Chang}},
  \bibinfo {author} {\bibfnamefont {Alberto}\ \bibnamefont
  {Fern{\'a}ndez-Nieves}}, \bibinfo {author} {\bibfnamefont {Seth}\
  \bibnamefont {Fraden}}, \ and\ \bibinfo {author} {\bibfnamefont {Zvonimir}\
  \bibnamefont {Dogic}},\ }\bibfield  {title} {\enquote {\bibinfo {title}
  {Transition from turbulent to coherent flows in confined three-dimensional
  active fluids},}\ }\href
  {http://science.sciencemag.org/content/355/6331/eaal1979} {\bibfield
  {journal} {\bibinfo  {journal} {Science}\ }\textbf {\bibinfo {volume} {355}}
  (\bibinfo {year} {2017})}\BibitemShut {NoStop}%
\bibitem [{\citenamefont {Zhang}\ \emph {et~al.}(2017)\citenamefont {Zhang},
  \citenamefont {Luijten}, \citenamefont {Grzybowski},\ and\ \citenamefont
  {Granick}}]{Granick_Review}%
  \BibitemOpen
  \bibfield  {author} {\bibinfo {author} {\bibfnamefont {Jie}\ \bibnamefont
  {Zhang}}, \bibinfo {author} {\bibfnamefont {Erik}\ \bibnamefont {Luijten}},
  \bibinfo {author} {\bibfnamefont {Bartosz~A.}\ \bibnamefont {Grzybowski}}, \
  and\ \bibinfo {author} {\bibfnamefont {Steve}\ \bibnamefont {Granick}},\
  }\bibfield  {title} {\enquote {\bibinfo {title} {Active colloids with
  collective mobility status and research opportunities},}\ }\href
  {http://dx.doi.org/10.1039/C7CS00461C} {\bibfield  {journal} {\bibinfo
  {journal} {Chem. Soc. Rev.}\ }\textbf {\bibinfo {volume} {46}},\ \bibinfo
  {pages} {5551--5569} (\bibinfo {year} {2017})}\BibitemShut {NoStop}%
\bibitem [{\citenamefont {Marchetti}\ \emph {et~al.}(2013)\citenamefont
  {Marchetti}, \citenamefont {Joanny}, \citenamefont {Ramaswamy}, \citenamefont
  {Liverpool}, \citenamefont {Prost}, \citenamefont {Rao},\ and\ \citenamefont
  {Simha}}]{Marchetti_Review}%
  \BibitemOpen
  \bibfield  {author} {\bibinfo {author} {\bibfnamefont {M.~C.}\ \bibnamefont
  {Marchetti}}, \bibinfo {author} {\bibfnamefont {J.~F.}\ \bibnamefont
  {Joanny}}, \bibinfo {author} {\bibfnamefont {S.}~\bibnamefont {Ramaswamy}},
  \bibinfo {author} {\bibfnamefont {T.~B.}\ \bibnamefont {Liverpool}}, \bibinfo
  {author} {\bibfnamefont {J.}~\bibnamefont {Prost}}, \bibinfo {author}
  {\bibfnamefont {Madan}\ \bibnamefont {Rao}}, \ and\ \bibinfo {author}
  {\bibfnamefont {R.~Aditi}\ \bibnamefont {Simha}},\ }\bibfield  {title}
  {\enquote {\bibinfo {title} {Hydrodynamics of soft active matter},}\ }\href
  {\doibase 10.1103/RevModPhys.85.1143} {\bibfield  {journal} {\bibinfo
  {journal} {Rev. Mod. Phys.}\ }\textbf {\bibinfo {volume} {85}},\ \bibinfo
  {pages} {1143--1189} (\bibinfo {year} {2013})}\BibitemShut {NoStop}%
\bibitem [{\citenamefont {Vicsek}\ and\ \citenamefont
  {Zafeiris}(2012)}]{Vicsek_Review}%
  \BibitemOpen
  \bibfield  {author} {\bibinfo {author} {\bibfnamefont {Tam\'as}\ \bibnamefont
  {Vicsek}}\ and\ \bibinfo {author} {\bibfnamefont {Anna}\ \bibnamefont
  {Zafeiris}},\ }\bibfield  {title} {\enquote {\bibinfo {title} {Collective
  motion},}\ }\href
  {http://www.sciencedirect.com/science/article/pii/S0370157312000968}
  {\bibfield  {journal} {\bibinfo  {journal} {Physics Reports}\ }\textbf
  {\bibinfo {volume} {517}},\ \bibinfo {pages} {71 -- 140} (\bibinfo {year}
  {2012})}\BibitemShut {NoStop}%
\bibitem [{\citenamefont {Toner}\ \emph {et~al.}(2005)\citenamefont {Toner},
  \citenamefont {Tu},\ and\ \citenamefont {Ramaswamy}}]{Toner_Review}%
  \BibitemOpen
  \bibfield  {author} {\bibinfo {author} {\bibfnamefont {John}\ \bibnamefont
  {Toner}}, \bibinfo {author} {\bibfnamefont {Yuhai}\ \bibnamefont {Tu}}, \
  and\ \bibinfo {author} {\bibfnamefont {Sriram}\ \bibnamefont {Ramaswamy}},\
  }\bibfield  {title} {\enquote {\bibinfo {title} {Hydrodynamics and phases of
  flocks},}\ }\href {\doibase https://doi.org/10.1016/j.aop.2005.04.011}
  {\bibfield  {journal} {\bibinfo  {journal} {Annals of Physics}\ }\textbf
  {\bibinfo {volume} {318}},\ \bibinfo {pages} {170 -- 244} (\bibinfo {year}
  {2005})},\ \bibinfo {note} {special Issue}\BibitemShut {NoStop}%
\bibitem [{\citenamefont {Czir\'ok}\ \emph {et~al.}(1997)\citenamefont
  {Czir\'ok}, \citenamefont {Stanley},\ and\ \citenamefont
  {Vicsek}}]{Vicsek1997}%
  \BibitemOpen
  \bibfield  {author} {\bibinfo {author} {\bibfnamefont {Andr\'as}\
  \bibnamefont {Czir\'ok}}, \bibinfo {author} {\bibfnamefont {H.~Eugene}\
  \bibnamefont {Stanley}}, \ and\ \bibinfo {author} {\bibfnamefont {Tam\'as}\
  \bibnamefont {Vicsek}},\ }\bibfield  {title} {\enquote {\bibinfo {title}
  {Spontaneously ordered motion of self-propelled particles},}\ }\href
  {http://stacks.iop.org/0305-4470/30/i=5/a=009} {\bibfield  {journal}
  {\bibinfo  {journal} {Journal of Physics A: Mathematical and General}\
  }\textbf {\bibinfo {volume} {30}},\ \bibinfo {pages} {1375} (\bibinfo {year}
  {1997})}\BibitemShut {NoStop}%
\bibitem [{\citenamefont {Kyriakopoulos}\ \emph {et~al.}(2016)\citenamefont
  {Kyriakopoulos}, \citenamefont {Ginelli},\ and\ \citenamefont
  {Toner}}]{Toner2016}%
  \BibitemOpen
  \bibfield  {author} {\bibinfo {author} {\bibfnamefont {Nikos}\ \bibnamefont
  {Kyriakopoulos}}, \bibinfo {author} {\bibfnamefont {Francesco}\ \bibnamefont
  {Ginelli}}, \ and\ \bibinfo {author} {\bibfnamefont {John}\ \bibnamefont
  {Toner}},\ }\bibfield  {title} {\enquote {\bibinfo {title} {Leading birds by
  their beaks: the response of flocks to external perturbations},}\ }\href
  {http://iopscience.iop.org/article/10.1088/1367-2630/18/7/073039/meta}
  {\bibfield  {journal} {\bibinfo  {journal} {New Journal of Physics}\ }\textbf
  {\bibinfo {volume} {18}},\ \bibinfo {pages} {073039} (\bibinfo {year}
  {2016})}\BibitemShut {NoStop}%
\bibitem [{\citenamefont {Bricard}\ \emph {et~al.}(2015)\citenamefont
  {Bricard}, \citenamefont {Caussin}, \citenamefont {Das}, \citenamefont
  {Savoie}, \citenamefont {Chikkadi}, \citenamefont {Shitara}, \citenamefont
  {Chepizhko}, \citenamefont {Peruani}, \citenamefont {Saintillan},\ and\
  \citenamefont {Bartolo}}]{Bricard2015}%
  \BibitemOpen
  \bibfield  {author} {\bibinfo {author} {\bibfnamefont {Antoine}\ \bibnamefont
  {Bricard}}, \bibinfo {author} {\bibfnamefont {Jean-Baptiste}\ \bibnamefont
  {Caussin}}, \bibinfo {author} {\bibfnamefont {Debasish}\ \bibnamefont {Das}},
  \bibinfo {author} {\bibfnamefont {Charles}\ \bibnamefont {Savoie}}, \bibinfo
  {author} {\bibfnamefont {Vijayakumar}\ \bibnamefont {Chikkadi}}, \bibinfo
  {author} {\bibfnamefont {Kyohei}\ \bibnamefont {Shitara}}, \bibinfo {author}
  {\bibfnamefont {Oleksandr}\ \bibnamefont {Chepizhko}}, \bibinfo {author}
  {\bibfnamefont {Fernando}\ \bibnamefont {Peruani}}, \bibinfo {author}
  {\bibfnamefont {David}\ \bibnamefont {Saintillan}}, \ and\ \bibinfo {author}
  {\bibfnamefont {Denis}\ \bibnamefont {Bartolo}},\ }\bibfield  {title}
  {\enquote {\bibinfo {title} {Emergent vortices in populations of colloidal
  rollers},}\ }\href {https://www.nature.com/articles/ncomms8470} {\bibfield
  {journal} {\bibinfo  {journal} {Nature communications}\ }\textbf {\bibinfo
  {volume} {6}} (\bibinfo {year} {2015})}\BibitemShut {NoStop}%
\bibitem [{\citenamefont {Morin}\ \emph
  {et~al.}(2017{\natexlab{a}})\citenamefont {Morin}, \citenamefont
  {Desreumaux}, \citenamefont {Caussin},\ and\ \citenamefont
  {Bartolo}}]{Morin2017}%
  \BibitemOpen
  \bibfield  {author} {\bibinfo {author} {\bibfnamefont {Alexandre}\
  \bibnamefont {Morin}}, \bibinfo {author} {\bibfnamefont {Nicolas}\
  \bibnamefont {Desreumaux}}, \bibinfo {author} {\bibfnamefont {Jean-Baptiste}\
  \bibnamefont {Caussin}}, \ and\ \bibinfo {author} {\bibfnamefont {Denis}\
  \bibnamefont {Bartolo}},\ }\bibfield  {title} {\enquote {\bibinfo {title}
  {{Distortion and destruction of colloidal flocks in disordered
  environments}},}\ }\href {\doibase 10.1038/nphys3903} {\bibfield  {journal}
  {\bibinfo  {journal} {Nature Physics}\ }\textbf {\bibinfo {volume} {13}},\
  \bibinfo {pages} {63--67} (\bibinfo {year} {2017}{\natexlab{a}})}\BibitemShut
  {NoStop}%
\bibitem [{\citenamefont {Morin}\ \emph
  {et~al.}(2017{\natexlab{b}})\citenamefont {Morin}, \citenamefont {{Lopes
  Cardozo}}, \citenamefont {Chikkadi},\ and\ \citenamefont
  {Bartolo}}]{Morin2017b}%
  \BibitemOpen
  \bibfield  {author} {\bibinfo {author} {\bibfnamefont {Alexandre}\
  \bibnamefont {Morin}}, \bibinfo {author} {\bibfnamefont {David}\ \bibnamefont
  {{Lopes Cardozo}}}, \bibinfo {author} {\bibfnamefont {Vijayakumar}\
  \bibnamefont {Chikkadi}}, \ and\ \bibinfo {author} {\bibfnamefont {Denis}\
  \bibnamefont {Bartolo}},\ }\bibfield  {title} {\enquote {\bibinfo {title}
  {{Diffusion, subdiffusion, and localization of active colloids in random post
  lattices}},}\ }\href {\doibase 10.1103/PhysRevE.96.042611} {\bibfield
  {journal} {\bibinfo  {journal} {Physical Review E}\ }\textbf {\bibinfo
  {volume} {96}},\ \bibinfo {pages} {042611} (\bibinfo {year}
  {2017}{\natexlab{b}})}\BibitemShut {NoStop}%
\bibitem [{\citenamefont {Palacci}\ \emph {et~al.}(2015)\citenamefont
  {Palacci}, \citenamefont {Sacanna}, \citenamefont {Abramian}, \citenamefont
  {Barral}, \citenamefont {Hanson}, \citenamefont {Grosberg}, \citenamefont
  {Pine},\ and\ \citenamefont {Chaikin}}]{Palacci2015}%
  \BibitemOpen
  \bibfield  {author} {\bibinfo {author} {\bibfnamefont {J{\'e}r{\'e}mie}\
  \bibnamefont {Palacci}}, \bibinfo {author} {\bibfnamefont {Stefano}\
  \bibnamefont {Sacanna}}, \bibinfo {author} {\bibfnamefont {Ana{\"\i}s}\
  \bibnamefont {Abramian}}, \bibinfo {author} {\bibfnamefont {J{\'e}r{\'e}mie}\
  \bibnamefont {Barral}}, \bibinfo {author} {\bibfnamefont {Kasey}\
  \bibnamefont {Hanson}}, \bibinfo {author} {\bibfnamefont {Alexander~Y.}\
  \bibnamefont {Grosberg}}, \bibinfo {author} {\bibfnamefont {David~J.}\
  \bibnamefont {Pine}}, \ and\ \bibinfo {author} {\bibfnamefont {Paul~M.}\
  \bibnamefont {Chaikin}},\ }\bibfield  {title} {\enquote {\bibinfo {title}
  {Artificial rheotaxis},}\ }\href
  {http://advances.sciencemag.org/content/1/4/e1400214} {\bibfield  {journal}
  {\bibinfo  {journal} {Science Advances}\ }\textbf {\bibinfo {volume} {1}}
  (\bibinfo {year} {2015})}\BibitemShut {NoStop}%
\bibitem [{\citenamefont {Geyer}\ \emph {et~al.}(2018)\citenamefont {Geyer},
  \citenamefont {Morin},\ and\ \citenamefont {Bartolo}}]{Geyer2018}%
  \BibitemOpen
  \bibfield  {author} {\bibinfo {author} {\bibfnamefont {Delphine}\
  \bibnamefont {Geyer}}, \bibinfo {author} {\bibfnamefont {Alexandre}\
  \bibnamefont {Morin}}, \ and\ \bibinfo {author} {\bibfnamefont {Denis}\
  \bibnamefont {Bartolo}},\ }\href@noop {} {\enquote {\bibinfo {title} {{Sounds
  and hydrodynamics of polar active fluids}},}\ }\bibinfo {howpublished}
  {submitted} (\bibinfo {year} {2018})\BibitemShut {NoStop}%
\bibitem [{\citenamefont {Toner}\ and\ \citenamefont {Tu}(1995)}]{Toner95}%
  \BibitemOpen
  \bibfield  {author} {\bibinfo {author} {\bibfnamefont {John}\ \bibnamefont
  {Toner}}\ and\ \bibinfo {author} {\bibfnamefont {Yuhai}\ \bibnamefont {Tu}},\
  }\bibfield  {title} {\enquote {\bibinfo {title} {Long-range order in a
  two-dimensional dynamical $\mathrm{XY}$ model: How birds fly together},}\
  }\href {\doibase 10.1103/PhysRevLett.75.4326} {\bibfield  {journal} {\bibinfo
   {journal} {Phys. Rev. Lett.}\ }\textbf {\bibinfo {volume} {75}},\ \bibinfo
  {pages} {4326--4329} (\bibinfo {year} {1995})}\BibitemShut {NoStop}%
\bibitem [{\citenamefont {Woodhouse}\ \emph {et~al.}(2016)\citenamefont
  {Woodhouse}, \citenamefont {Forrow}, \citenamefont {Fawcett},\ and\
  \citenamefont {Dunkel}}]{DunkelPNAS}%
  \BibitemOpen
  \bibfield  {author} {\bibinfo {author} {\bibfnamefont {Francis~G.}\
  \bibnamefont {Woodhouse}}, \bibinfo {author} {\bibfnamefont {Aden}\
  \bibnamefont {Forrow}}, \bibinfo {author} {\bibfnamefont {Joanna~B.}\
  \bibnamefont {Fawcett}}, \ and\ \bibinfo {author} {\bibfnamefont
  {J{\"{o}}rn}\ \bibnamefont {Dunkel}},\ }\bibfield  {title} {\enquote
  {\bibinfo {title} {{Stochastic cycle selection in active flow networks}},}\
  }\href {\doibase 10.1073/pnas.1603351113} {\bibfield  {journal} {\bibinfo
  {journal} {Proceedings of the National Academy of Sciences}\ }\textbf
  {\bibinfo {volume} {113}},\ \bibinfo {pages} {8200--8205} (\bibinfo {year}
  {2016})},\ \Eprint {http://arxiv.org/abs/1607.08015} {arXiv:1607.08015}
  \BibitemShut {NoStop}%
\bibitem [{\citenamefont {Forrow}\ \emph {et~al.}(2017)\citenamefont {Forrow},
  \citenamefont {Woodhouse},\ and\ \citenamefont {Dunkel}}]{DunkelPRL}%
  \BibitemOpen
  \bibfield  {author} {\bibinfo {author} {\bibfnamefont {Aden}\ \bibnamefont
  {Forrow}}, \bibinfo {author} {\bibfnamefont {Francis~G.}\ \bibnamefont
  {Woodhouse}}, \ and\ \bibinfo {author} {\bibfnamefont {J{\"{o}}rn}\
  \bibnamefont {Dunkel}},\ }\bibfield  {title} {\enquote {\bibinfo {title}
  {{Mode selection in compressible active flow networks}},}\ }\href {\doibase
  10.1103/PhysRevLett.119.028102} {\bibfield  {journal} {\bibinfo  {journal}
  {Physical Review Letters}\ }\textbf {\bibinfo {volume} {119}},\ \bibinfo
  {pages} {1--6} (\bibinfo {year} {2017})}\BibitemShut {NoStop}%
\bibitem [{\citenamefont {Woodhouse}\ and\ \citenamefont
  {Dunkel}(2017)}]{DunkelNatComm}%
  \BibitemOpen
  \bibfield  {author} {\bibinfo {author} {\bibfnamefont {Francis~G.}\
  \bibnamefont {Woodhouse}}\ and\ \bibinfo {author} {\bibfnamefont
  {J{\"{o}}rn}\ \bibnamefont {Dunkel}},\ }\bibfield  {title} {\enquote
  {\bibinfo {title} {{Active matter logic for autonomous microfluidics}},}\
  }\href {\doibase 10.1038/ncomms15169} {\bibfield  {journal} {\bibinfo
  {journal} {Nature Communications}\ }\textbf {\bibinfo {volume} {8}},\
  \bibinfo {pages} {15169} (\bibinfo {year} {2017})},\ \Eprint
  {http://arxiv.org/abs/1610.05515} {arXiv:1610.05515} \BibitemShut {NoStop}%
\bibitem [{\citenamefont {Nekorkin}(2015)}]{Nekorkin2015}%
  \BibitemOpen
  \bibfield  {author} {\bibinfo {author} {\bibfnamefont {Vladimir~I.}\
  \bibnamefont {Nekorkin}},\ }\href
  {http://doi.wiley.com/10.1002/9783527695942} {\emph {\bibinfo {title}
  {{Introduction to nonlinear oscillations}}}}\ (\bibinfo  {publisher}
  {Wiley-VCH Verlag GmbH {\&} Co. KGaA},\ \bibinfo {address} {Weinheim,
  Germany},\ \bibinfo {year} {2015})\BibitemShut {NoStop}%
\bibitem [{\citenamefont {Quincke}(1896)}]{Quincke}%
  \BibitemOpen
  \bibfield  {author} {\bibinfo {author} {\bibfnamefont {G.}~\bibnamefont
  {Quincke}},\ }\bibfield  {title} {\enquote {\bibinfo {title} {Uber rotationen
  im constanten electrischen felde},}\ }\href
  {http://onlinelibrary.wiley.com/doi/10.1002/andp.18962951102/abstract}
  {\bibfield  {journal} {\bibinfo  {journal} {Annalen der Physik}\ ,\ \bibinfo
  {pages} {417--486}} (\bibinfo {year} {1896})}\BibitemShut {NoStop}%
\bibitem [{\citenamefont {Melcher}\ and\ \citenamefont
  {Taylor}(1969)}]{Taylor69}%
  \BibitemOpen
  \bibfield  {author} {\bibinfo {author} {\bibfnamefont {J.~R.}\ \bibnamefont
  {Melcher}}\ and\ \bibinfo {author} {\bibfnamefont {G.~I.}\ \bibnamefont
  {Taylor}},\ }\bibfield  {title} {\enquote {\bibinfo {title}
  {Electrohydrodynamics: A review of the role of interfacial shear stresses},}\
  }\href {\doibase 10.1146/annurev.fl.01.010169.000551} {\bibfield  {journal}
  {\bibinfo  {journal} {Annual Review of Fluid Mechanics}\ }\textbf {\bibinfo
  {volume} {1}},\ \bibinfo {pages} {111--146} (\bibinfo {year}
  {1969})}\BibitemShut {NoStop}%
\bibitem [{\citenamefont {Crocker}\ and\ \citenamefont {Grier}(1996)}]{Grier}%
  \BibitemOpen
  \bibfield  {author} {\bibinfo {author} {\bibfnamefont {John~C.}\ \bibnamefont
  {Crocker}}\ and\ \bibinfo {author} {\bibfnamefont {David~G.}\ \bibnamefont
  {Grier}},\ }\bibfield  {title} {\enquote {\bibinfo {title} {Methods of
  digital video microscopy for colloidal studies},}\ }\href
  {http://www.sciencedirect.com/science/article/pii/S0021979796902179}
  {\bibfield  {journal} {\bibinfo  {journal} {Journal of colloid and interface
  science}\ }\textbf {\bibinfo {volume} {179}},\ \bibinfo {pages} {298--310}
  (\bibinfo {year} {1996})}\BibitemShut {NoStop}%
\bibitem [{\citenamefont {Lu}\ \emph {et~al.}(2007)\citenamefont {Lu},
  \citenamefont {Sims}, \citenamefont {Oki}, \citenamefont {Macarthur},\ and\
  \citenamefont {Weitz}}]{Lu2007}%
  \BibitemOpen
  \bibfield  {author} {\bibinfo {author} {\bibfnamefont {Peter~J.}\
  \bibnamefont {Lu}}, \bibinfo {author} {\bibfnamefont {Peter~A.}\ \bibnamefont
  {Sims}}, \bibinfo {author} {\bibfnamefont {Hidekazu}\ \bibnamefont {Oki}},
  \bibinfo {author} {\bibfnamefont {James~B.}\ \bibnamefont {Macarthur}}, \
  and\ \bibinfo {author} {\bibfnamefont {David~A.}\ \bibnamefont {Weitz}},\
  }\bibfield  {title} {\enquote {\bibinfo {title} {{Target-locking acquisition
  with real-time confocal (TARC) microscopy}},}\ }\href {\doibase
  10.1364/OE.15.008702} {\bibfield  {journal} {\bibinfo  {journal} {Optics
  Express}\ }\textbf {\bibinfo {volume} {15}},\ \bibinfo {pages} {8702}
  (\bibinfo {year} {2007})}\BibitemShut {NoStop}%
\bibitem [{\citenamefont {Blair}\ and\ \citenamefont {Dufresne}()}]{Blair}%
  \BibitemOpen
  \bibfield  {author} {\bibinfo {author} {\bibfnamefont {Daniel}\ \bibnamefont
  {Blair}}\ and\ \bibinfo {author} {\bibfnamefont {Eric}\ \bibnamefont
  {Dufresne}},\ }\href@noop {} {\enquote {\bibinfo {title} {The matlab particle
  tracking code repository},}\ }\bibinfo {howpublished} {retrieved from
  \url{http://physics.georgetown .edu/matlab/}}\BibitemShut {NoStop}%
\bibitem [{Note1()}]{Note1}%
  \BibitemOpen
  \bibinfo {note} {We focus on the situation where $\alpha >D_{\protect \rm
  T}q^2$. In the case of extreme confinement where this condition is not met,
  flow reversal can occur only via local melting.}\BibitemShut {Stop}%
\end{thebibliography}

%

\begin{figure*}
\begin{center}
\includegraphics[width=2\columnwidth]{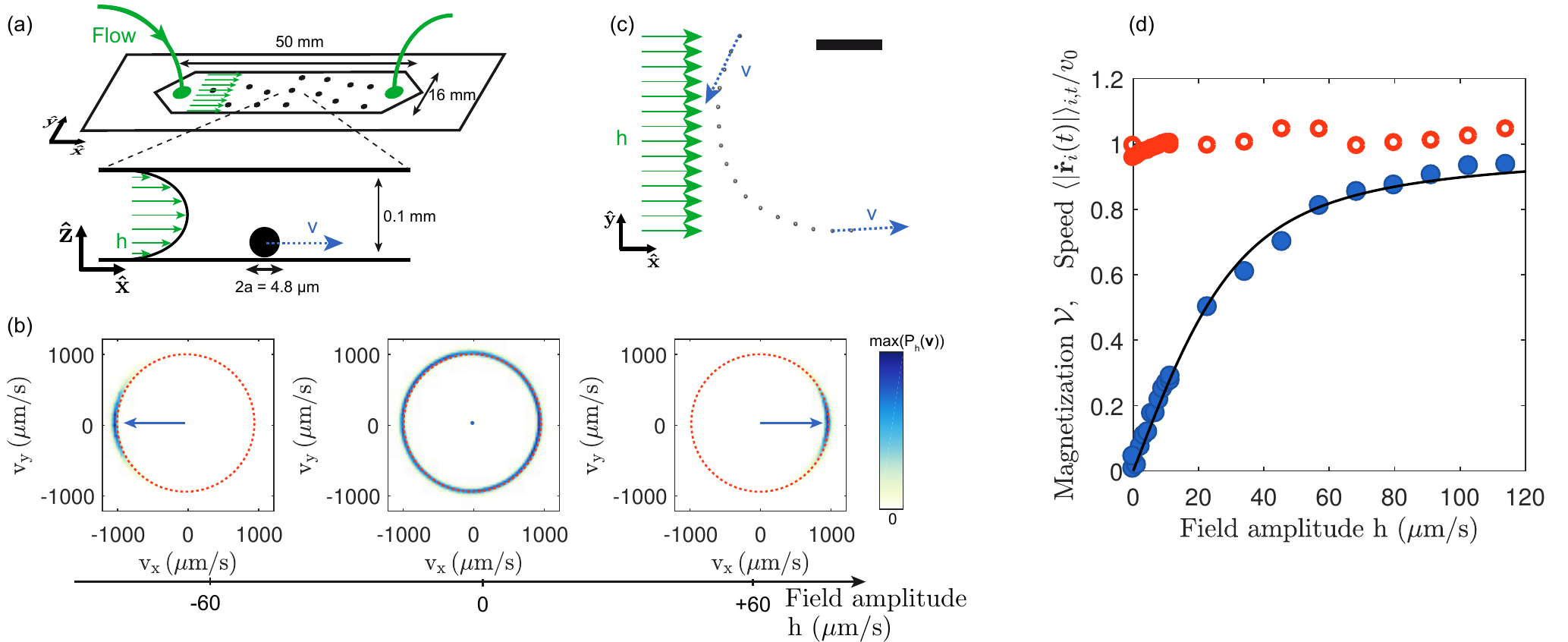}
\caption{{Isolated colloidal rollers align with  external flows as XY spins with magnetic fields.}
(a) Sketch of the microfluidic setup. Top: top view. Bottom: side view. Colloidal rollers (black dots) of diameter $2a=4.8\,\rm \mu m$ and speed $\mathbf{v}$ (blue arrow) are confined in a $50\,\rm mm\times 16\,mm\times 0.1\,mm$ channel. Hexadecane is injected through the channel at a constant flow rate. A Poiseuille flow results in the $\mathbf{\hat{z}}$ direction (green arrows). The strength of the external field $h$ is defined as the magnitude of the hexadecane flow at a distance $z=a$ from the bottom wall.
(b) Probability density function of the roller velocities for three different hexadecane flows $\rm P_h(\mathbf{v})$. The dashed circles are guidelines corresponding to $v=v_0$. When $h=0$, the distribution is isotropic and strongly peaked around $v=v_0$. When $h\neq0$, the distributions are biased along the direction of $\mathbf{h}$, resulting in finite average velocities (blue arrows).
(c) Superimposed images of a colloidal roller in a hexadecane flow ($h=+60\,\rm\mu m/s$). Time interval between two pictures: 30~ms. The roller reorients at constant speed along the flow direction. Scale bar: $100\,\rm\mu m$.
(d) Blue circles: Magnetization curve $\mathcal V(h)$ of the noninteracting rollers. Red open circles: time and ensemble average of the normalized roller speed  $\langle |\mathbf{\dot{r}_i}(t)| \rangle_{i,t} / v_0$.  The rotational mobility $\mu=0.08\,\rm\mu m^{-1}$ is evaluated from the best fit of the magnetization curve with the theoretical formula (dark solid line): ${\mathcal V}=I_1(\mu h/D)/I_0(\mu h/D)$, where  $I_n$ is the modified Bessel function of order $n$.  
\label{Fig1}
}
\end{center}
\end{figure*}
\begin{figure*}
\begin{center}
\includegraphics[width=2\columnwidth]{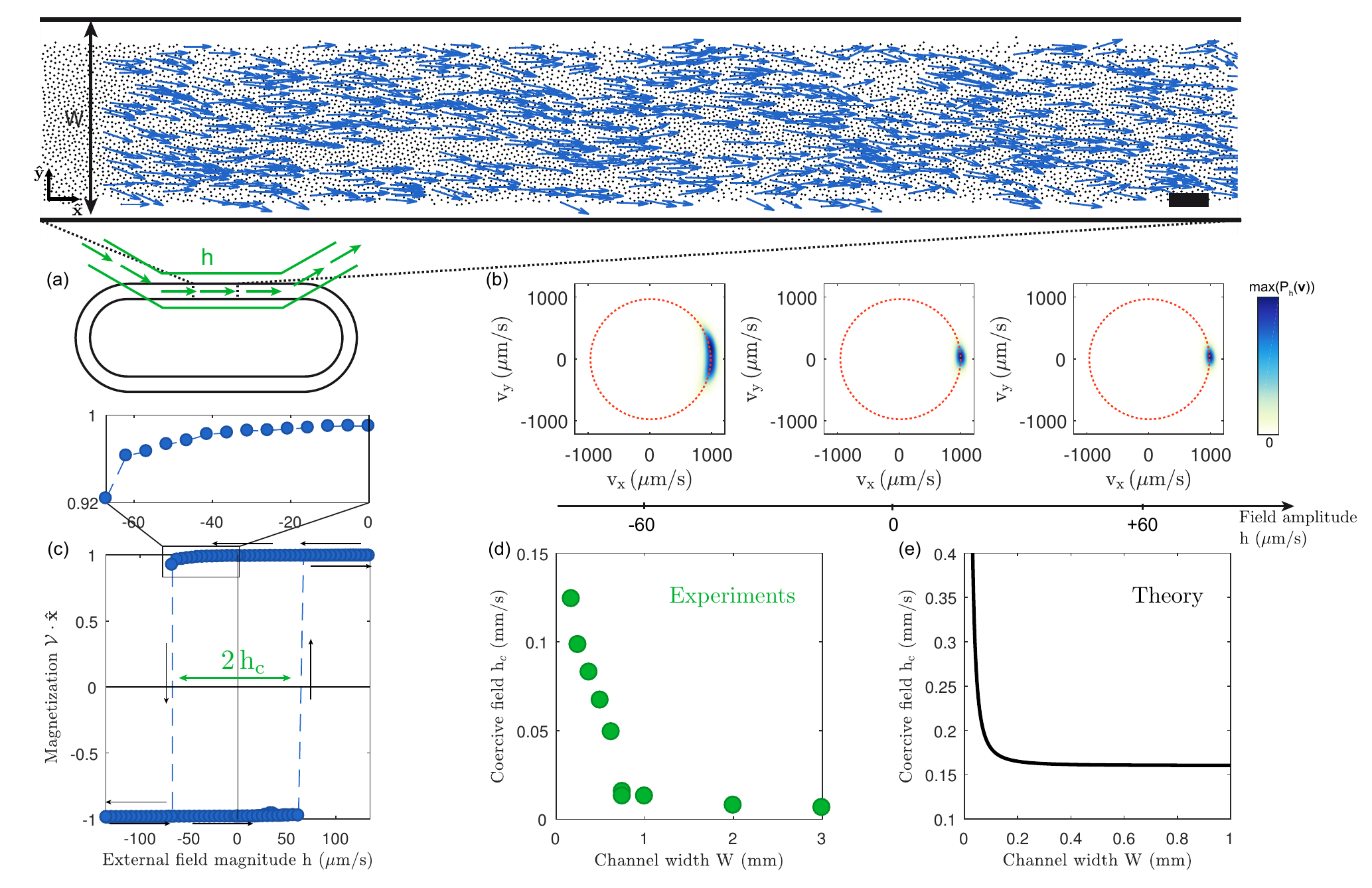}
\caption{{Non-linear response of polar liquids to longitudinal external fields.}
(a) Sketch of the microfluidic geometry. The rollers are confined in a photolithographied race-track (black). A constant hexadecane flow is applied to the rollers by means of an additional microfluidic channel (green). Top picture: close-up on the homogeneous polar liquid flowing in a race-track of width $W=0.5\,\rm mm$ when $h=0$. In this experiment, the polar liquid spontaneously flows along the clockwise direction as indicated by the instantaneous particle velocities (the velocity of one roller out of ten is plotted, blue arrows). Scale bar: $100\,\rm \mu m$.  
(b) Probability density function of the roller velocities in the polar-liquid state. At $h=0$, the distribution is biased revealing the spontaneous symmetry breaking of the roller orientation. At $h=+60\,\rm \mu m/s$ transverse velocity fluctuations are reduced. At $h=-60\,\rm \mu m/s$, the polar liquid cruises against the external field and velocity fluctuations are enhanced. The speed of the rollers $v_0$ (red dotted circles) is left unchanged by the external field.
(c) Magnetization curve $\VV\cdot\hat{\mathbf x} (\mathbf{h})$ of the polar liquid. Upon cycling the external field $\mathbf{h}$ (dark arrows) the active ferromagnet displays an hysteretic behavior. The coercive field $h_{\rm c}$ is defined as the width of the hysteresis loop. Top panel: close-up showing the minute decrease of the magnetization prior to reversal.
{\color{black}(d) The coercive field $h_{\rm c}$  decreases with the channel width $W$ (green dots). (e)  Theoretical prediction  $h_{\rm c}(W)=(2/\sqrt{243})\alpha v_0\left(1+2D_{\rm T} q^2/\alpha\right)^{3/2}$.  As a technical remark, we note that the limit  $W\to\infty$  is not relevant. Our two-mode model is intrinsically based on the scale separation between the variations of the longitudinal and transverse components of  $\mathbf v$: $v_x$ varies over the scale of the channel length  while $v_y$ varies over the channel width. Unconfined polar liquids cannot resist to external fields.}
\label{Fig2}
}
\end{center}
\end{figure*}
\begin{figure*}
\begin{center}
\includegraphics[width=2\columnwidth]{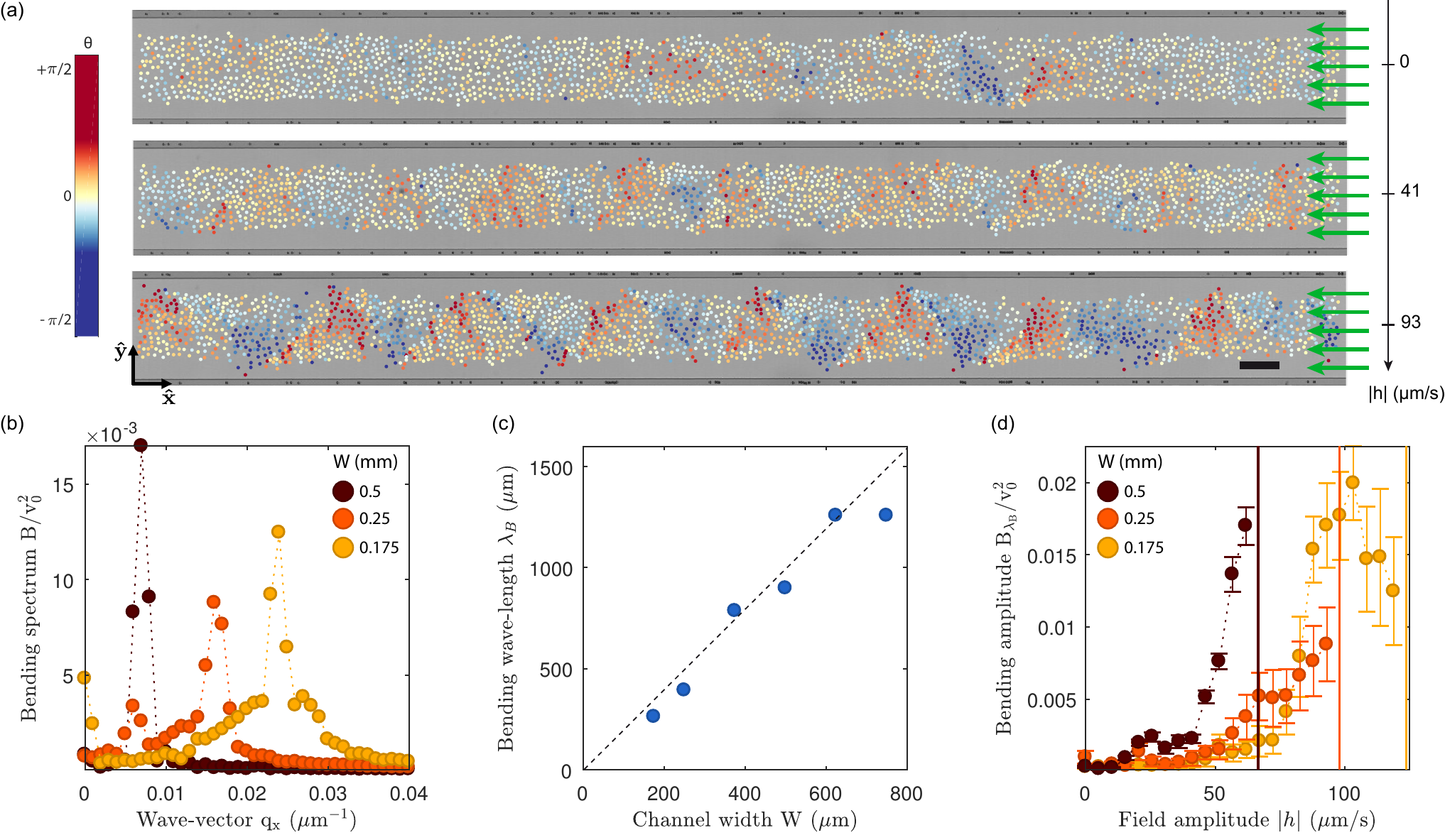}
\caption{{Bending deformations of a polar liquid flowing against an external field.}
(a) Snapshots of polar liquids flowing against external fields of increasing magnitude. The color indicates  the angle of the instantaneous velocity of the rollers $\theta$. As $|h|$ increases, a bending-oscillation pattern grows. $W=0.25\,\rm mm$. Scale bar: $100\,\rm \mu m$. 
(b) The time-average spectra of the bending modes along $\mathbf{q} = (q_x,0)$ ($B(q_x)=\langle |v_y(q_x,t)|^2\rangle_t$) are plotted for three different confinement widths $W=0.5,\,0.25,\,0.175\,\rm mm$. The values of $h$ are taken just before flow reversal: $h= 62,\,93,\,119\,\rm \mu m/s$, respectively. $\lambda_{\rm B}$ is defined by the wavelength where $B$ is maximal.
(c) Variations of $\lambda_{\rm B}$ with $W$. Dashed line: $\lambda_{\rm B} = 2W$.
(d) Variations of the amplitude of the bending mode at $q=2\pi/\lambda_{\rm B}$ with $|h|$. The bending deformations at $\lambda_{\rm B}$ increase when increasing $|h|$.  The vertical lines indicate the values of $h_{\rm c}(W)$ for the three experiments. Error bars: 1 sd (time statistics).
\label{Fig3}
}
\end{center}
\end{figure*}
\begin{figure*}
\begin{center}
\includegraphics[width=2\columnwidth]{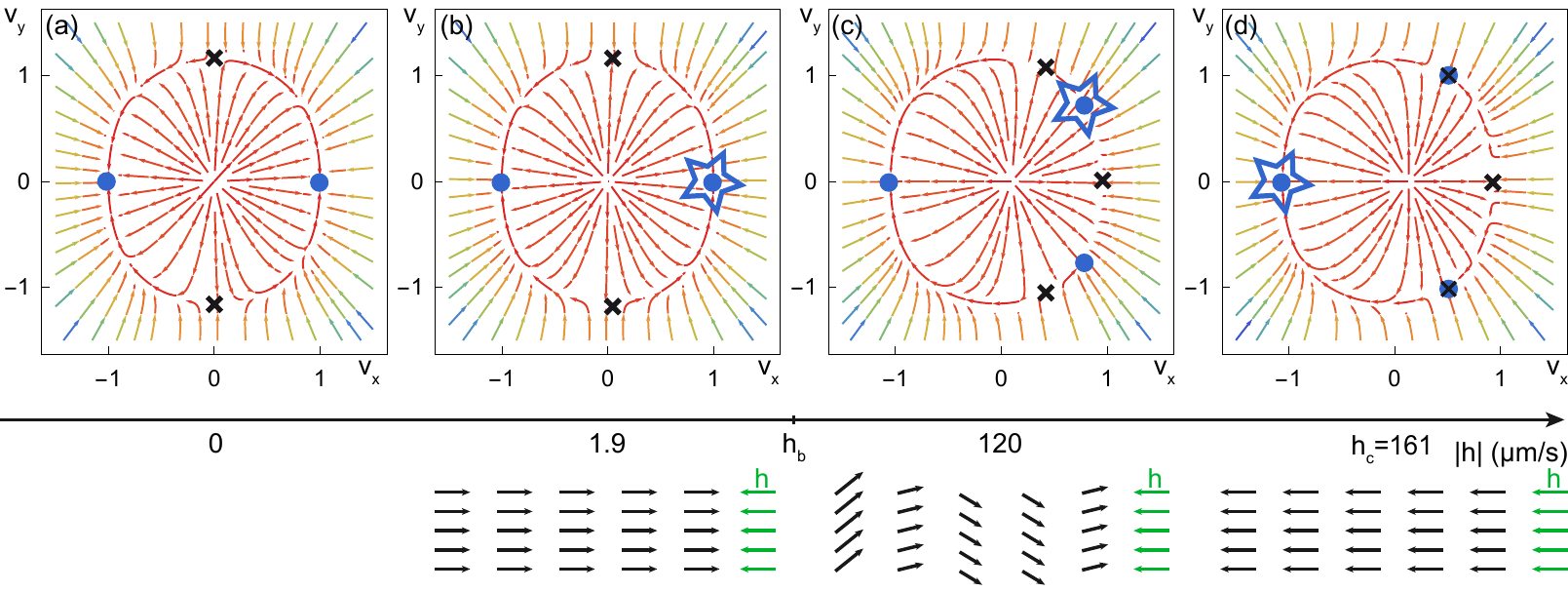}
\caption{{Two-mode theory of the hysteretic response.}
We look for stationary solutions of the hydrodynamics of polar liquids taking the form $\mathbf{v} = v_x(t)\mathbf{\hat{x}} + v_y(t)\cos(qx-\omega t)\mathbf{\hat{y}}$. The problem then reduces to studying a dynamical system: $\partial_t \mathbf{V} = \mathbf{F}(\mathbf{V},h)$ where $\mathbf{V} =(v_x,v_y)$. 
(a) Force-field $\mathbf{F}(\mathbf{V},h=0)$ in the absence of external field, {plotted for $W=0.5\,\rm mm$}.  The blue circles  correspond to stable fixed points where $F=0$. The crosses indicate the position of saddle points. A trivial unstable point is located at $(0,0)$, not shown.
(b) Force-field $\mathbf{F}(\mathbf{V},h)$ at finite $h$. The star symbol indicates the stationary position of the dynamical system. When $|h|<h_{\rm B}$, the system stays in the stable fixed point corresponding to a uniform longitudinal flow in the direction opposite to $\mathbf h$ as sketched in the bottom panel.
(c) Increasing $|h|$ the homogeneous solution of (b) becomes unstable. Another stable point emerges between two saddle points and corresponds to the buckled flow sketched in the bottom panel. Such stable buckled flows are consistent with the experimental observations of Fig.~\ref{Fig3}.
(d) At $|h|=h_{\rm c}$  the topmost saddle point collides with the stable fixed point (superimposed cross and dot symbols). As a result, the only stable conformation corresponds to a flow aligned with $\mathbf h$. 
\label{Fig4}
}
\end{center}
\end{figure*}
\begin{figure*}
\begin{center}
\includegraphics[width=2\columnwidth]{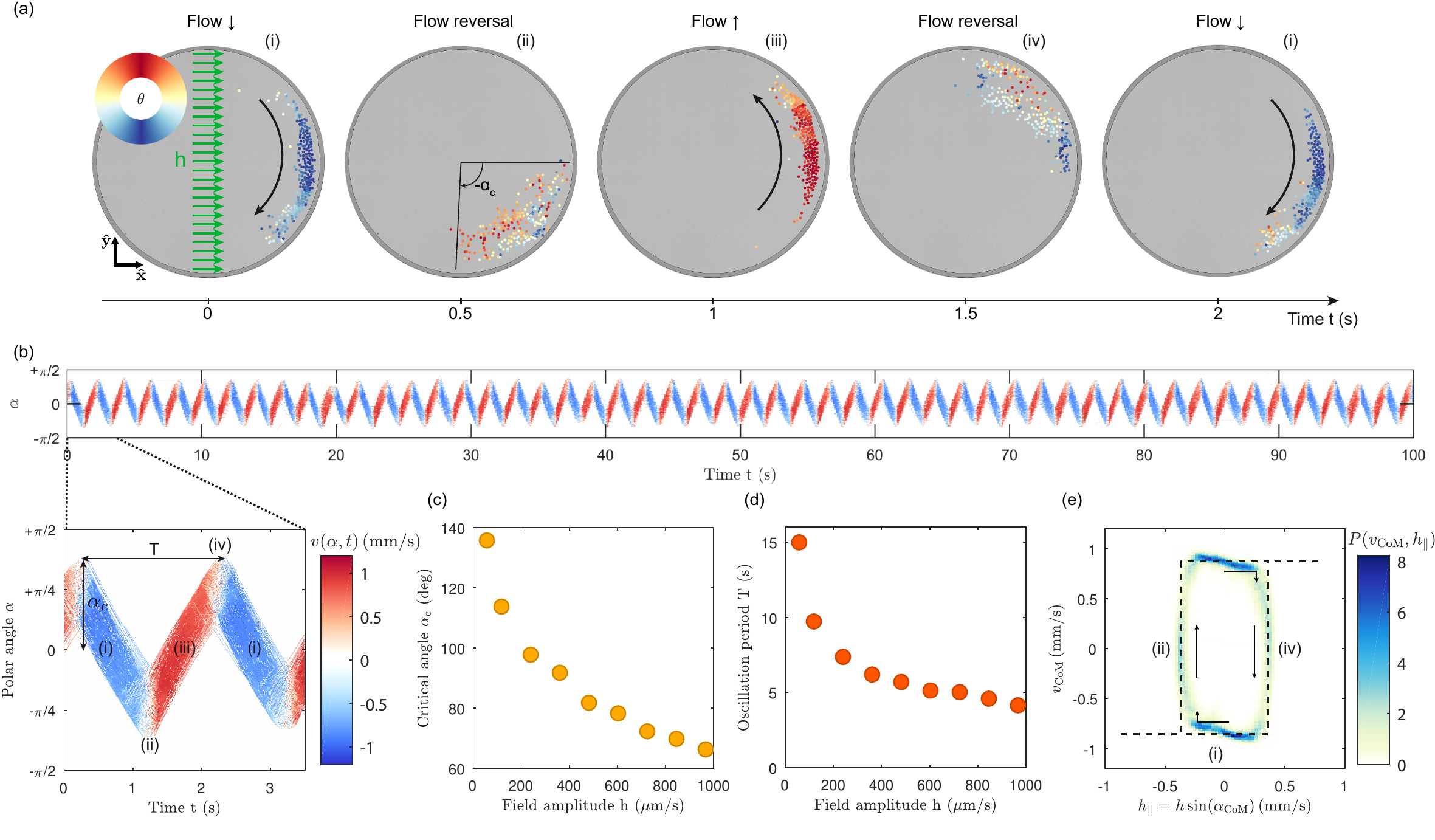}
\caption{{Spontaneous oscillations of a polar-liquid droplet.}
(a) Subsequent pictures of a polar liquid droplet oscillating spontaneously in a circular chamber of radius $R=0.5\,\rm mm$. The green arrows indicate the  direction of $\mathbf h$. The color of the particles indicates the instantaneous direction of their velocity $\theta$. The polar-liquid droplet reverses its motion when it reaches a critical angle $\pm\alpha_{\rm c}$ along the curved boundary.  $h=485\,\rm \mu m/s$.
(b) Time variations of the radial average of the active-liquid flow $v(\alpha,t)$ showing a well defined period $T$ and amplitude $\alpha_{\rm c}$ of oscillations. The four states defined in (a) are indicated in the close-up view.
(c) The critical angle $\alpha_c$ is reduced when increasing the magnitude of the external field  $h$. $R=2.3\,\rm mm$.
(d) The period of the oscillations $T$ is reduced when increasing the magnitude of the external field  $h$. $R=2.3\,\rm mm$.
(e) Density of probability $P(v_{\rm CoM},h_{\parallel})$. The support of this probability is defined by the hysteresis cycle. Dashed line: sketch of the dynamical response of the polar liquid. Arrows: direction of the cycle exploration. The four states defined in (a) correspond to the four branches of the hysteresis loop.  
\label{Fig5}
}
\end{center}
\end{figure*}
\begin{figure*}
\begin{center}
\includegraphics[width=2\columnwidth]{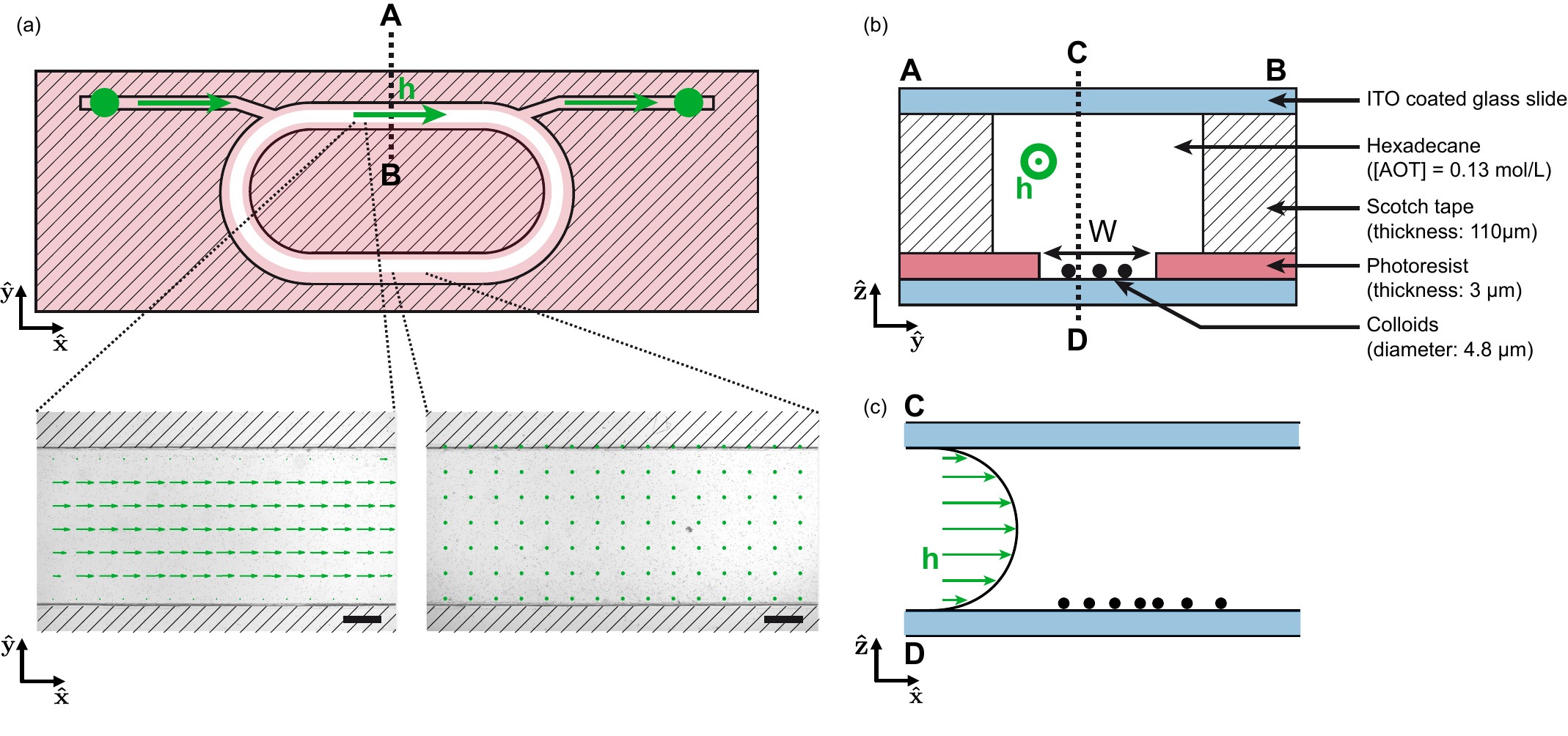}
\caption{{\color{black}{Schematics of the microfluidic device.}
(a) Top view. The photoresist pattern is shown in red color. The racetrack along which the active-liquid flows corresponds to the white path. The geometry of the fluidic channel corresponds to the black solid lines. The flow direction is indicated with a green arrow.  The two bottom pictures show the solvent flow field measured by particle image velocimetry seeding the hexadecane oil with colloids. Due to the imbalance between the two hydrodynamic resistances, the residual flow in the bottom branch is negligible (right picture) compared to the flow in the observation window (left picture). Scale bar: 500 $\mu$m
(b) Side view in the A-B section defined in (a). Two ITO coated glass slides are assembled with a double-sided tape. The colloids roll on the bottom ITO surface and are confined by the photoresist pattern (red color).
 (c) Side view in the C-D section as defined in (b). The hexadecane flow has a Poiseuille profile along the $z$-direction.}
\label{Fig:experience}
}
\end{center}
\end{figure*}
\begin{figure*}
\begin{center}
\includegraphics[width=\columnwidth]{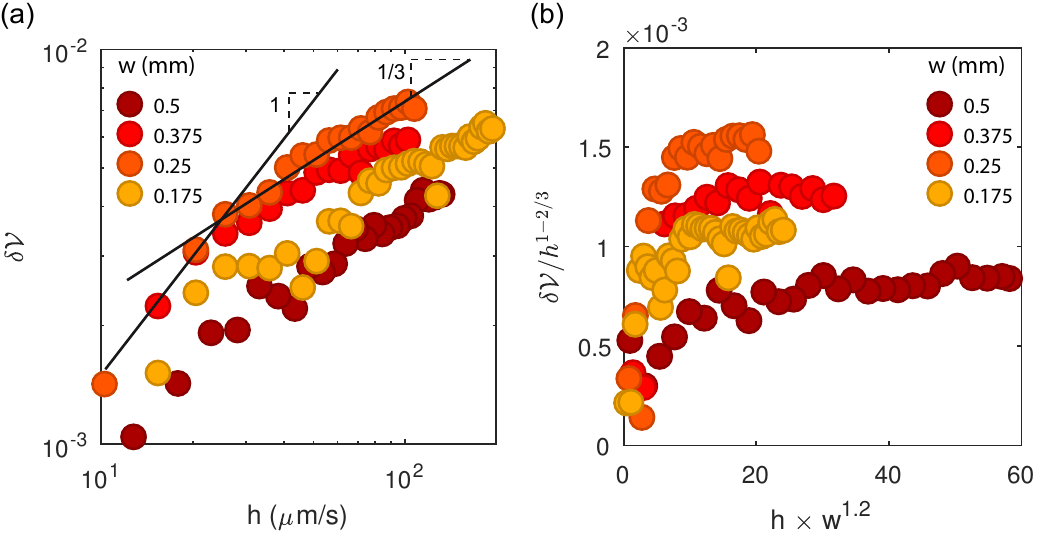}
\caption{{Non-linear response of a polar liquid flowing in the same direction as the external field.}
(a) The increase of $\delta {\mathcal V}\equiv |{\mathcal V}(h)-{\mathcal V}(h=0)|$ is linear at small $h$ and consistent with a $h^{1/3}$ scaling at high $h$. This behavior is compatible with the theoretical  predictions of~\cite{Toner2016} based on renormalization group calculations. 
(b) Finite-size scaling of $\delta {\mathcal V}(h,W)$. Unlike what would be expected from the RG theory of~\cite{Toner2016}, all plots do not collapse on the same master curve.
\label{FigSupp1}
}
\end{center}
\end{figure*}

\end{document}